\documentclass[reprint,amsmath,amssymb,pra,twocolumn,longbibliography]{revtex4-2}
\usepackage{amsmath}
\usepackage{mathrsfs}
 \usepackage{graphicx}
 \usepackage{csquotes}
\usepackage{color}
\usepackage{ulem}

\def\ptl{ IEEE Photon.\ Technol.\ Lett.}

\def\ol{Opt. Lett.}

\def\pra{Phys. Rev. A}
\def\prl{Phys. Rev. Lett.}

\begin{document}
\title{On the robustness of bound states in the continuum in  \\
  waveguides   with lateral leakage channels}  

\author{Lijun Yuan}
\email{Corresponding author: ljyuan@ctbu.edu.cn}
\affiliation{College of Mathematics and Statistics, Chongqing Technology and Business University, Chongqing, 
China \\ Chongqing Key Laboratory of Social Economic and Applied Statistics, Chongqing Technology and Business University, Chongqing, China }
\author{Ya Yan Lu}
\affiliation{Department of Mathematics, City University of Hong Kong, Hong Kong}

\begin{abstract}
Bound states in the continuum (BICs) are  trapped or guided modes
with frequencies in radiation continua. They are associated with
high-quality-factor resonances that give rise to strong local field
enhancement and rapid 
variations in scattering spectra, and have found many valuable  
applications. A guided mode of an optical waveguide can also be a BIC,
if there is a lateral structure supporting compatible waves 
propagating in the lateral direction, i.e., there is a 
channel for lateral leakage. A BIC is typically destroyed
(becomes a resonant or a leaky mode) if the structure is slightly
perturbed, but some BICs are robust with respect to a large family of
perturbations. In this paper, we show (analytically and
numerically)  that a typical BIC in optical
waveguides with a left-right mirror symmetry and a single lateral
leakage channel is robust with respect to any structural perturbation that
preserves the left-right mirror symmetry. Our study improves the
theoretical understanding on BICs and can be useful when applications
of BICs in optical waveguides are explored. 
\end{abstract}
\maketitle

\section{Introduction}

A bound state in the continuum (BIC) is a trapped or guided mode that
decays to zero in a spatial direction along which radiation modes with
the same frequency and wavevector (when appropriate) can propagate to
or from infinity \cite{neumann29,hsu16,kosh19}. In photonics, BICs
have been found on various structures including waveguides with a local
distortion \cite{evans94}, periodic structures sandwiched between two
homogeneous media \cite{mari08,lee12,hsu13_2,bulg14b},
waveguides with lateral leakage channels
\cite{bonnet97,plot11,weim13,zou15,hope16,bezus18,nguyen19,yu19a,yu19,byk20,yu20},
rotationally symmetric periodic structures surrounded by a homogeneous medium \cite{bulg17prl,sadbel19}, and
anisotropic multilayer structures \cite{gomis17}. 
Applications of the BICs are mostly related to
resonances with arbitrarily large quality factors ($Q$ factors) that
appear when the structure or the wavevector are properly
perturbed \cite{kosh18,lijun18pra,hu18,lijun20,zhen20b}. High-$Q$
resonances give rise to strong local field 
enhancement and abrupt variations in transmission/reflection spectra
that are useful for lasing, sensing, switching, and nonlinear optics
applications.

An important theoretical question is about the robustness (i.e., the
continual existence) of the BICs with respect to small structural
perturbations. If a BIC is protected by a symmetry, i.e., there is a
symmetry mismatch between the BIC and the radiation modes, it is
typically robust with respect to symmetry-preserving
perturbations. For periodic structures with the  in-plane inversion
symmetry and the up-down mirror symmetry, some BICs unprotected by
symmetry are also robust with respect to 
perturbations preserving these two symmetries
\cite{zhen14,bulg17pra,yuan17_4,yuan20b,conrob}. An optical waveguide
may have a lateral leakage channel, if the structure away from the
waveguide core has guided modes that can propagate in the lateral
direction. A BIC on such a structure may or may
not be symmetry-protected. In a recent work \cite{byk20}, Bykov {\it
  et al.} studied a waveguide with a rectangular core in a background
with a slab, showed that even the BICs unprotected by symmetry are 
robust with respect to variations in the structural parameters.

In this paper, we analyze the robustness of BICs in optical waveguides
with respect to arbitrary structural perturbations. Based on an
all-order perturbation method first developed in \cite{yuan17_4}, we
show analytically that a generic BIC in a waveguide with a lateral
mirror symmetry and a single radiation channel is robust with respect 
to any perturbation that preserves the lateral mirror symmetry. The
rest of this paper is organized as follows. In Sect.~II, we present
the necessary background concerning the waveguide structure and the
BICs. In Sect.~III, we formulate a related scattering problem and
present some useful properties of the scattering solutions. Section IV
contains the main result about robustness, including the precise
conditions and the proof. In Sect.~V, we present numerical
examples to illustrate the continuous existence of BICs when
structural parameters are varied. The paper is concluded  with a brief
discussion in Sect.~VI.

\section{Guided modes and BICs}

\begin{figure}[http]
 \centering 
 \includegraphics[scale=0.5]{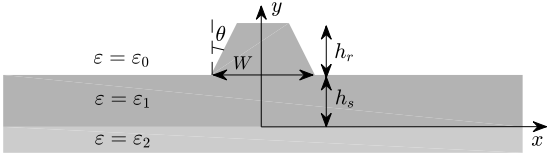}
 \caption{A waveguide with an isosceles trapezoidal ridge of height $h_r$, base  $W$ and base angle $\pi/2-\theta$.
   The thickness of the slab is $h_s$. The dielectric constants of the
   slab and the ridge, the substrate, and the cladding are
   $\varepsilon_1$, $\varepsilon_2$, and  $\varepsilon_0$ satisfying
   $\varepsilon_1 > \varepsilon_2 \ge \varepsilon_0\ge 1$.} 
 \label{fig:structure}
 \end{figure}

We consider a three-dimensional (3D) $z$-invariant lossless dielectric
waveguide with a core embedded in a layered background. The dielectric function
$\varepsilon$ of the waveguide depends only on the two
transverse variables $x$ and $y$, that is,
$\varepsilon=\varepsilon({\bf r})$ for ${\bf r}=(x,y)$.  We assume
the structure is symmetric in $x$, the core is 
contained in the region given by $|x| < W/2$ for some $W>0$, and the
background is a layered medium with a dielectric function
$\varepsilon_s$ that depends only on $y$. Thus 
\begin{eqnarray}
  && \label{eq:Eps_symm}
\varepsilon({\bf r}) = \varepsilon(-x,y), \quad \forall \ {\bf r}  \in \mathbb{R}^2, \\
  &&\label{hasslab}
     \varepsilon({\bf r}) = \varepsilon_s(y), \quad \mbox{if}  \ |x| > W/2.
\end{eqnarray}
In addition, we assume the background itself is a 2D waveguide,
typically consisting of a slab, a substrate and a cladding. We denote the dielectric
constants of the cladding and the substrate by $\varepsilon_0$ and
$\varepsilon_2$, respectively, and assume $\varepsilon_2 \ge
\varepsilon_0 \ge 1$.
As an example, we show a ridge waveguide in Fig.~\ref{fig:structure}, 
where the ridge is an isosceles trapezoid. 


For such a 3D $z$-invariant waveguide, a guided mode decays to zero as
$r=\sqrt{x^2 +  y^2} \to \infty$, and depends on $z$ as $\exp ( i \gamma
  z)$,  where $\gamma$ is a real constant (the propagating constant). The electric
  field of the guided mode can be written as $\mbox{Re} [ {\bf E}\,
  e^{i   (\gamma z - \omega t)} ]$, where $\omega$ is the angular frequency,
and ${\bf E}$ is a vector function that depends only on
$\mathbf{r}$. From the frequency-domain Maxwell's equations, it is
easy to show 
that ${\bf E}$ satisfies   
  \begin{eqnarray}
 \label{eq:MaxEq1}
&& \left( \nabla + i \gamma \mathbf{e}_3 \right)  \times  \left(
   \nabla + i \gamma \mathbf{e}_3 \right)  \times \mathbf{E} - k^2
   \varepsilon({\bf r}) \mathbf{E} = 0, \\
&& \label{eq:MaxEq2}
  \left( \nabla + i \gamma \mathbf{e}_3 \right) \cdot
   [ \varepsilon({\bf r}) \mathbf{E} ] = 0,
  \end{eqnarray}
  where $k = \omega / c$ is the freespace wavenumber, $c$ is the speed
  of light in vacuum,   and $\mathbf{e}_3 = (0, 0, 1)$ is the unit
  vector in the $z$ direction. For simplicity, we call ${\bf E}$
  the electric field and normalize the guided mode such that 
\begin{equation}
\label{eq:NormalizationBIC}
\frac{1}{L^2} \int\limits_{\mathbb{R}^2} \varepsilon(\mathbf{r}) |
\mathbf{E}(\mathbf{r}) |^2 \,  d \mathbf{r} = 1,
\end{equation}
where $L$ is a characteristic length.


  Since $\varepsilon$ is real, it is easy to verify that the vector field 
\begin{equation}
\label{tildeE} 
\tilde{\bf E}({\bf r}) =
\left[ \begin{matrix}  \overline{E}_x({\bf 
      r}) \\ \overline{E}_y({\bf r}) \\ -\overline{E}_z({\bf 
      r}) \end{matrix} \right], 
\end{equation}
also satisfies Eqs.~(\ref{eq:MaxEq1}) and (\ref{eq:MaxEq2}),
where  $E_x$, $E_y$ and $E_z$ are the components of ${\bf E}$, 
$\overline{E}_x$ is the complex conjugate of $E_x$, etc. If the
guided mode is non-degenerate, then there is a constant  
$C_0$ such that $\mathbf{E}({\bf r}) = C_0 \tilde{\bf E}({\bf r})$. 
Since the power carried by the guided mode  is finite, we 
must have $|C_0|=1$. If $C_0 = e^{ i \varphi_0}$ for a real phase $\varphi_0$, we 
can replace ${\bf E}({\bf r})$ by $e^{-i \varphi_0/2} \mathbf{E}({\bf 
  r})$ and obtain ${\bf E}({\bf r}) = \tilde{\bf E}({\bf r})$. 
Therefore, the guided mode can be scaled such that
\begin{equation}
  \label{eq:RealCondition}
\left[ \begin{matrix}  E_x({\bf r}) \\ E_y({\bf r}) \\ E_z({\bf 
      r}) \end{matrix} \right]
  =
\left[ \begin{matrix}  \overline{E}_x({\bf 
      r}) \\ \overline{E}_y({\bf r}) \\ -\overline{E}_z({\bf 
      r}) \end{matrix} \right].
\end{equation}
That is, the $x$ and $y$ components of ${\bf E}$ are real and the $z$
component of ${\bf E}$ is pure imaginary. 
  

Since the structure has a reflection symmetry in $x$, the field
  components of a guided mode is either even in $x$ or odd in
  $x$. More precisely, let ${\bf E}({\bf r})$ be the electric field of a guided 
  mode and $\hat{\bf E}({\bf r})$ be the vector field given by 
  \begin{equation}
    \label{hatE}
    \hat{\bf E}({\bf r}) = \left[
      \begin{matrix}
        E_x(-x,y) \cr -E_y(-x, y) \cr -E_z(-x,y) 
      \end{matrix} \right],  
  \end{equation}
    then ${\bf E}$,   $\hat{\bf E}$, $  ( {\bf E}+\hat{\bf E} )/2$ and
  $ ( {\bf E} -\hat{\bf E} )/2$ all satisfy   Eqs.~(\ref{eq:MaxEq1}) and
  (\ref{eq:MaxEq2}).  Since ${\bf E}$ can be replaced by $  ( {\bf
    E}+\hat{\bf E} )/2$
  or   $ ( {\bf E} -\hat{\bf E} )/2$, we can assume either ${\bf E} =
  \hat{\bf E}$, i.e.,
  \begin{equation}
\label{eq:EvenInX}
\left[ \begin{matrix}  E_x({\bf r}) \\ E_y({\bf r}) \\ E_z({\bf 
      r}) \end{matrix} \right]
= \left[ \begin{matrix}
    E_x(-x,y) \\  -E_y(-x,y) \\ -E_z(-x,y) 
    \end{matrix} \right], 
  \end{equation}
  or
  ${\bf E} = -\hat{\bf E}$, i.e.,
  \begin{equation}
     \label{eq:OddInX}
\left[ \begin{matrix}  E_x({\bf r}) \\ E_y({\bf r}) \\ E_z({\bf 
      r}) \end{matrix} \right]
= \left[ \begin{matrix}
    -E_x(-x,y) \\  E_y(-x,y) \\ E_z(-x,y) 
    \end{matrix} \right]. 
  \end{equation}

 
  The layered background given by the dielectric function
  $\varepsilon_s(y)$ is a 2D waveguide with transverse-electronic (TE)
  and transverse-magnetic (TM) modes. To distinguish these modes from
  the eigenmodes of the original 3D waveguide,  we call them slab modes. A TE
  slab mode is characterized by a zero $y$ component of the electric
  field and a scalar function $u(y)$ satisfying the  eigenvalue problem: 
\begin{equation}
\label{eq:1DHelm_TE}
\frac{d^2 u}{dy^2}  + k^2 \varepsilon_s(y) u = k^2 ( \eta^{\rm te} )^2
u,  \quad -\infty < y < +\infty, 
\end{equation}
with the condition $u(y) \to 0$ as $y \to \pm \infty$, where
$\eta^{\rm te}$ is the real positive effective index of the slab
mode ($k\eta^{\rm te}$ is the propagation constant). 
  Since the slab mode must decay exponentially in the 
substrate and the cladding, we have 
$\eta^{\rm te} > \sqrt{\varepsilon_2}$. In addition, we assume the TE
slab mode  is normalized such that 
\[
  \frac{1}{L} \int_{-\infty}^{+\infty} |u(y)|^2 dy = 1,
\]
for the same $L$ in Eq.~(\ref{eq:NormalizationBIC}).
In the region given by $x > W/2$, if this TE slab mode also depends on
$z$ as $\exp( i \gamma z)$ and is outgoing or tends to zero as $x \to +\infty$,  then its electric field is given by
\begin{equation}
\label{eq:TEp}
\mathbf{E}({\bf r}) = {\bf f}_{+}^{\rm te}(y) e^{ i   \alpha^{\rm te}  x},
\end{equation}
where 
\begin{eqnarray}
  &&   \label{alpte} 
     \alpha^{\rm te} = \sqrt{  (k  \eta^{\rm te} ) ^2 - \gamma^2}, \\
  && \label{Fpte}
     {\bf  f}_{+}^{\rm te}(y) = \dfrac{1}{k \eta^{\rm te}}  \left[  \begin{matrix} - 
    \gamma \\ 0 \\ \alpha^{\rm te}  \end{matrix} \right]  u(y).    
\end{eqnarray}
Notice that if $ \gamma < k \eta^{\rm te}$, then $\alpha^{\rm te} > 0$
and the field given in Eq.~(\ref{eq:TEp}) is outgoing as $x \to
+\infty$. If $\gamma > k \eta^{\rm te}$, then $\alpha^{\rm te}$ is pure
imaginary with a positive imaginary part, and the field decays to zero
as $x \to +\infty$. For $x < -W/2$, the TE slab mode with the
same $z$ dependence has an electric field given by 
\begin{equation}
\label{eq:TEn}
\mathbf{E}({\bf r}) = {\bf f}_{-}^{\rm te}(y) e^{ - i \alpha^{\rm te} x},
\end{equation}
where
\begin{equation}
  \label{Fmte}
{\bf f}_{-}^{\rm te}(y) = \dfrac{-1}{k \eta^{\rm te}}
\left[  \begin{matrix}  \gamma \\ 0 \\ \alpha^{\rm te}   \end{matrix}
\right]  u(y), 
\end{equation}
and it is either outgoing or tends to zero as $x \to -\infty$.


The case for the TM slab mode is similar. It has a zero $y$ component in the magnetic  field and a scalar function $v(y)$ satisfying the eigenvalue problem
\begin{equation}
\label{eq:1DHelm_TM}
\dfrac{d }{d y} \left[ \dfrac{1}{\varepsilon_s(y)} \dfrac{d v}{d y}
\right] + k^2 v = \frac{ (k \eta^{\rm tm} )^2 }{\varepsilon_s(y)} v, \quad
-\infty < y < \infty,  
\end{equation}
with the condition $v(y) \to 0$ as $y \to \pm \infty$, where
$\eta^{\rm tm} > \sqrt{\varepsilon_2} > 0$ is the effective index of
  the mode. Similar to the TE modes, we normalize $v(y)$ such that
\[
  \frac{1}{L}\int_{-\infty}^{+\infty} \frac{1}{\varepsilon_s(y)}
  |v(y)|^2 \, dy =   1. 
\]
Assuming the same dependence on $z$, the electric field of the TM slab mode can be
written down as  
\begin{eqnarray}
  \label{eq:TMp}
&&  \mathbf{E}({\bf r}) = {\bf f}_{+}^{\rm tm}(y) e^{ i \alpha^{\rm tm} x}, \ \ 
  x > \frac{W}{2}, \\
&&   
  \label{eq:TMn}  
   \mathbf{E}({\bf r})= {\bf f}_{-}^{\rm tm}(y) e^{ - i \alpha^{\rm tm}
   x},   \ \   x < -\frac{W}{2},  
\end{eqnarray}
where
\begin{eqnarray}
  && \label{alptm}
     \alpha^{\rm tm} = \sqrt{ (k \eta^{\rm tm})^2 - \gamma^2}, \\
  && \label{Fptm} {\bf f}_{+}^{\rm tm}(y)=  \dfrac{1} { k \eta^{\rm tm} \varepsilon_s(y)}
  \left[  \begin{matrix} i \alpha^{\rm tm}  v'(y) \\  (k \eta^{\rm 
        tm})^2  v(y) \\  i \gamma v'(y)   \end{matrix} \right],    \\
&& \label{Fmtm} {\bf f}_{-}^{\rm tm}(y) =   \dfrac{-1}{k \eta^{\rm tm} \varepsilon_s(y)}
  \left[  \begin{matrix} -i \alpha^{\rm tm}  v'(y) \\  (k \eta^{\rm 
        tm})^2  v(y) \\  i \gamma v'(y)   \end{matrix} \right], 
\end{eqnarray}
and $v'(y) = dv/dy$. The TM slab mode with its electric field given in Eqs.~(\ref{eq:TMp}) and (\ref{eq:TMn})  is either outgoing (if $\gamma  < k \eta^{\rm tm}$) or evanescent (if $\gamma > k \eta^{\rm tm}$) as $x \to \pm \infty$.


At a fixed frequency, the layered background has a finite number of
slab modes. We order the TE and TM slab modes separately according to
their effective indices, and denote them as $\left\{ u_j, \eta_j^{\rm te}
\right\} $ for $j = 0$, 1, $\ldots$, $N-1$, and $\left\{
  v_j, \eta_j^{\rm tm} \right\} $ for $j = 0$, 1,  $\ldots$,
$M-1$, where $N$ and $M$ are the numbers of TE and TM modes,
respectively.  It is known that the TE and TM modes are
interlaced, namely, their effective indices satisfy 
\[
  \eta_0^{\rm te} > \eta_0^{\rm tm} > \eta_1^{\rm te} > \eta_1^{\rm
  tm} > \dots > \sqrt{\varepsilon_2}.
\]


In the slab region given by $|x| > W/2$,  a general time-harmonic field is a
superposition of the finite number of slab modes and a continuum of
radiation modes \cite{marcuse}. If the field depends on $z$ as $e^{i \gamma z}$ and
is outgoing as $x \to \pm \infty$, then the electric field can be written as
\begin{equation}
\label{eq:SlabExpE}
  {\bf E} = {\bf E}_{\pm}^{\rm te} + {\bf E}_{\pm,\diamond}^{\rm te} +
{\bf E}_{\pm}^{\rm tm} + {\bf E}_{\pm,\diamond}^{\rm tm}, 
\end{equation}
where the ``$+$" and ``$-$" subscripts are chosen for $x > W/2$
and  $x < -W/2$, respectively, and 
\begin{eqnarray}
&& \mathbf{E}_{\pm}^{\rm te}  =  \sum\limits_{j=0}^{N-1} a_j^{\pm}
  \mathbf{f}^{\rm te}_{\pm, j}(y) e^{  \pm  i \alpha^{\rm te}_j x},  \\
&& {\bf E}_{\pm,\diamond}^{\rm te} = \int_{-\infty}^{+\infty} a_\diamond^\pm(\beta)
   \mathbf{f}^{\rm te}_{\pm,\diamond}(y;\beta) e^{\pm i \alpha_\diamond (\beta) x} d
   \beta, \\
&& {\bf E}_{\pm}^{\rm tm} = \sum\limits_{j=0}^{M-1} b_j^{\pm}
   \mathbf{f}^{\rm tm}_{\pm, j}(y) e^{ \pm i \alpha^{\rm tm}_j x},    \\
&& {\bf E}_{\pm,\diamond}^{\rm tm} =    \int_{-\infty}^{+\infty} b^{\pm}_\diamond(\beta)
   \mathbf{f}^{\rm tm}_{\pm,\diamond}(y;\beta) e^{ \pm i \alpha_\diamond (\beta) x}
   d\beta. 
\end{eqnarray}
In the above,  $\alpha_j^{\rm te}$, $\alpha_j^{\rm
  tm}$, $\displaystyle 
{\bf f}_{+, j}^{\rm te}$, ${\bf f}_{-, j}^{\rm te}$, 
${\bf f}_{+, j}^{\rm tm}$, ${\bf f}_{-, j}^{\rm tm}$ are defined for the
$j$th TE or TM mode following Eqs.~(\ref{alpte}), (\ref{alptm}), (\ref{Fpte}), (\ref{Fmte}),
(\ref{Fptm}) and (\ref{Fmtm}); $a_j^\pm$ and $b_j^\pm$
are the coefficients of the slab modes; and the terms with the 
subscript ``$\diamond$'' are radiation modes that
depend continuously on a real wavenumber $\beta$ for the $y$ direction.
The wavenumber for the $x$ direction is 
\begin{equation}
  \label{radalp}
\alpha_\diamond(\beta) = \sqrt{k^2 \varepsilon_2 - \gamma^2 - \beta^2},   
\end{equation}
where $\varepsilon_2$ is the dielectric constant of the substrate. For
each $\beta$, ${\bf   f}_{\pm,\diamond}^{\rm te}$ and ${\bf  f}_{\pm,\diamond}^{\rm tm}$
are the TE and TM radiation modes \cite{marcuse}, and $a^\pm_\diamond$ and $b^\pm_\diamond$
are the corresponding coefficients.


If ${\bf E}$ is the electric field of a guided mode (of the 3D
waveguide) with a frequency $\omega$ and a propagation constant $\gamma$, then it
must satisfy Eq.~(\ref{eq:SlabExpE}) for properly chosen coefficients $\{
a_j^\pm, a_\diamond^\pm(\beta), b_j^\pm, b_\diamond^\pm(\beta) \}$. A regular
guided mode satisfies the condition
$k \eta_0^{\rm te}   < \gamma$, 
thus all $\alpha_j^{\rm te}$, $\alpha_j^{\rm tm}$ and
$\alpha_{\diamond}(\beta)$  are pure imaginary with positive imaginary parts,  and
${\bf E} \to 0$ as $x \to \pm \infty$.  A BIC is a special guided
mode with a positive propagation constant satisfying
  \begin{equation}
  \gamma < k \eta_0^{\rm te}.
  \end{equation}
This means that $\alpha_0^{\rm te}$ and probably
other $\alpha_j^{\rm tm}$ or $\alpha_j^{\rm te}$ are positive, thus
the layered background has at least one slab mode that can radiate
power to $x = \pm \infty$ and propagate in the $z$ direction as the
BIC. Since the BIC must decay to zero as $x \to \pm \infty$, it is
clear that $a_0^\pm$ and probably the coefficients of other slab modes (that
radiate out power in the $x$ direction) must be zero.
We are particularly interested in BICs  satisfying
\begin{equation}
\label{eq:OneGuidedMode}
 k \eta_0^{\rm tm} < \gamma <  k \eta_0^{\rm te}.
\end{equation}
In that case, $\alpha^{\rm te}_0$ is real, all other $\alpha^{\rm
  te}_j$, $\alpha^{\rm tm}_j$, and $\alpha_{\diamond}(\beta)$  are
pure imaginary. Therefore, only the fundamental TE slab mode can
radiate power to $x = \pm \infty$, and the coefficient $a_0^{\pm}$ of
the BIC must be zero.  Due to the reflection symmetry
  of the structure in the $x$ 
direction, we have either $a_0^+=  a_0^-$ or $a_0^+= - a_0^-$,  if the
BIC satisfies  (\ref{eq:EvenInX}) or (\ref{eq:OddInX}), respectively.

\section{Scattering solutions}
\label{sec:scattering}

For $\omega$ and $\gamma$ satisfying condition
(\ref{eq:OneGuidedMode}), the fundamental TE slab mode can
propagate in the slab region ($|x|> W/2$) with the $(x,z)$ dependence
given by $\exp [ i (\gamma z \pm  \alpha_0^{\rm te} x)]$, where $\alpha_0^{\rm
  te} = [ (k\eta_0^{\rm te})^2 - \gamma^2]^{1/2} > 0$.
This implies that we can consider scattering problems using the fundamental
TE slab mode as incident waves. The solutions of the scattering
problems are needed in Sect.~IV for proving the robustness of
BICs. 
If
\[
  {\bf u}_{+}({\bf r}) = {\bf f}_{+,0}^{\rm te}(y) \, e^{ i   \alpha_0^{\rm te} x}
\]
(with the assumed $z$ dependence $e^{i \gamma z}$) is the
incident wave in the left slab region ($x < -W/2$), then the total field contains a 
reflected wave $ R {\bf u}_{-}({\bf r})$, for
\[
  {\bf u}_{-}({\bf r})  =
  {\bf f}_{-,0}^{\rm te}(y) \, e^{  -i \alpha_0^{\rm te} x},
\]
in the left slab region, and a transmitted wave $T {\bf u}_{+}({\bf r})$ in the 
right slab region ($x > W/2$), where $R$ and $T$ are the reflection 
and transmission coefficients. Due to the reflection symmetry in $x$,
if we specify ${\bf u}_{-}({\bf r})$ as the incident wave in the right
slab region, the total field contains a reflected wave and a
transmitted wave in the right and left slab regions, respectively, and
the reflection and transmission coefficients are exactly the
same. In addition, if condition (\ref{eq:OneGuidedMode}) is satisfied, the
fundamental TE slab mode provides the only radiation channel, thus, 
the  $2 \times 2$ scattering matrix $\displaystyle
\left[ \begin{matrix} R & T \cr T & R \end{matrix} 
\right]$ is unitary. This implies that 
\begin{equation}
  |R|^2 + |T|^2 = 1, \ R\overline{T} + \overline{R} T = 0, \ |R\pm T|^2 = 1.  
\end{equation}

To construct a scattering solution satisfying the symmetry condition
(\ref{eq:EvenInX}) or (\ref{eq:OddInX}), we need to specify incident waves
in both left and right slab regions. For condition
(\ref{eq:EvenInX}), we let the incident waves be $C {\bf u}_+$ for $x < -W/2$ and $C
{\bf u}_{-}$ for $x > W/2$, where $C$ is a nonzero
constant, then any solution of this scattering problem satisfies 
\begin{equation}
\label{eq:ScatSolutionAsym1}
\mathbf{E}({\bf r}) \sim \left\{ \begin{matrix} {C \bf u}_{+}({\bf r}) 
     + C(R + T) \mathbf{u}_{-} ({\bf r}),  \quad x \to - \infty, \\
     C \mathbf{u}_{-}({\bf r})  + C (R + T) \mathbf{u}_{+} ({\bf r}), 
     \quad x \to  +\infty. \end{matrix} \right. 
\end{equation}
Although the far field given in the right hand side above already
satisfies condition (\ref{eq:EvenInX}), the total field ${\bf E}({\bf
  r})$ may not, because the solution is not unique if a  BIC exists at the same 
$\omega$ and $\gamma$.  
However, we can define a vector field
$\hat{\bf E}({\bf r})$ as in Eq.~(\ref{hatE}), then
$\hat{\bf E}$ and $({\bf E} + \hat{\bf E})/2$ solve the same
scattering problem. We can replace our original solution by $({\bf E}
+ \hat{\bf E})/2$, then the new solution, still denoted as ${\bf   E}({\bf
  r})$, satisfies condition (\ref{eq:EvenInX}). 

By choosing a proper constant $C$, we can ensure that condition
(\ref{eq:RealCondition})
is also satisfied.  Since $|R+T|=1$, we have
$R+T = e^{ i \varphi}$ for a real phase $\varphi$. If we let $C =
e^{-i\varphi/2}$, then Eq.~(\ref{eq:ScatSolutionAsym1}) becomes 
\begin{equation}
\label{getreal}
\mathbf{E}({\bf r}) \sim \left\{ \begin{matrix} {C \bf u}_{+}({\bf r}) 
     + \overline{C} \mathbf{u}_{-} ({\bf r}),  \quad x \to - \infty, \\
    C \mathbf{u}_{-}({\bf r})  + \overline{C} \mathbf{u}_{+} ({\bf r}), 
     \quad x \to  +\infty \end{matrix}. \right. 
\end{equation}
Due to the possible non-uniqueness, the above solution may not satisfy
condition (\ref{eq:RealCondition}).
However, we can define $\tilde{\bf E}$ as in
Eq.~(\ref{tildeE}), then
$\tilde{\bf E}$ and $({\bf E}+\tilde{\bf E})/2$ solve the same
scattering problem. If we replace the original solution by $({\bf
  E}+\tilde{\bf E})/2$, then the new  ${\bf E}({\bf r})$ satisfies
condition 
(\ref{eq:RealCondition}).

Since we are concerned with the case where a BIC exists at the same
$\omega$ and $\gamma$, we denote the electric fields of the BIC and the
scattering solution by ${\bf E}_0$ and ${\bf E}^{(s)}$,
respectively. Assuming both ${\bf E}_0$ and ${\bf   E}^{(s)}$ satisfy
conditions (\ref{eq:RealCondition}) and (\ref{eq:EvenInX}), we
can replace ${\bf E}^{(s)}$ by ${\bf E}^{(s)} + C_1 {\bf E}_0$ for a
real constant $C_1$ (if necessary), such that the following orthogonality
condition is satisfied:
\begin{equation}
  \label{orthogonal}
  \int_{\mathbb{R}^2}  \varepsilon({\bf r}) \overline{\bf E}^{(s)} \cdot {\bf
    E}_0\, d{\bf r} = 0. 
\end{equation}
In summary,  we have constructed a scattering
solution ${\bf E}^{(s)}$ that satisfies  (\ref{eq:RealCondition}),
(\ref{eq:EvenInX}) and (\ref{orthogonal}). 

Similarly, we can construct another scattering solution that satisfies 
conditions (\ref{eq:RealCondition}), (\ref{eq:OddInX}) and (\ref{orthogonal}).

\section{Robustness of BICs}
\label{sec:robust}

In this section, we study the robustness of BICs  in optical
waveguides based on an all-order perturbation method developed in our
previous works \cite{yuan17_4,conrob}. The robustness refers to the continual
existence of a BIC under small structural perturbations. The original unperturbed 
waveguide is given by a dielectric function
$\varepsilon_0({\bf   r})$ satisfying conditions (\ref{eq:Eps_symm})
and (\ref{hasslab}). It is assumed  that  the unperturbed waveguide has a
non-degenerate BIC with an electric field ${\bf   E}_0({\bf r})$, a
frequency $\omega_0$ and a propagation constant
$\gamma_0$. Moreover, $\omega_0$ and $\gamma_0$ must satisfy condition
(\ref{eq:OneGuidedMode}),  where $k$ should be replaced by
$k_0=\omega_0/c$, and $\eta_0^{\rm te}$ and $\eta_0^{\rm tm}$ are
effective indices of the fundamental TE and TM slab modes at frequency $\omega_0$. 
Without loss of generality, we assume $\mathbf{E}_0({\bf r})$ satisfies
conditions (\ref{eq:RealCondition}) and (\ref{eq:EvenInX}).
The dielectric function of the perturbed waveguide is given by 
\begin{equation}
\varepsilon({\bf r}) = \varepsilon_0({\bf r}) + \delta s({\bf r}),
\end{equation}
where $\delta$ is a small real number, $s({\bf r})$ is a real $O(1)$
function satisfying condition (\ref{eq:Eps_symm})  and $s({\bf r}) =
0$ if $|x| > W/2$  and if $|y|$ is sufficiently large. In the following,
we show that under a generic condition, the perturbed waveguide has a
BIC with a frequency $\omega$ near $\omega_0$, a propagation constant
$\gamma$ near $\gamma_0$, and an electric field ${\bf E}({\bf r})$ near
${\bf E}_0({\bf r})$.

Following the procedure developed in \cite{yuan17_4,conrob}, we construct
the BIC in the perturbed waveguide by  expanding $\mathbf{E}({\bf r})$,
$\gamma$ and $k = \omega/c$ in power series of $\delta$:
\begin{eqnarray}
\label{eq:expE}
&&  \mathbf{E}  = \mathbf{E}_0 +
                         \delta \mathbf{E}_1  + \delta^2
                         \mathbf{E}_2  + \ldots, \\ 
\label{eq:expGamma}
&& \gamma =  \gamma_0 +  \gamma_1 \delta + \gamma_2 \delta^2 + \ldots, \\
\label{eq:expk}
&& k = k_0 +  k_1 \delta + k_2 \delta^2 + \ldots.
\end{eqnarray}
Note that for any real $\gamma$ near $\gamma_0$, the governing equations
(\ref{eq:MaxEq1}) and (\ref{eq:MaxEq2})
have a solution satisfying
an outgoing radiation condition as $x \to \pm \infty$, but the
frequency $\omega$ is complex in general. Similarly, for any real
$\omega$ near $\omega_0$, there is an outgoing solution with a complex
$\gamma$. These outgoing 
solutions with a complex $\omega$ or a complex $\gamma$  are the resonant
and leaky modes, respectively. Our objective is to determine 
a BIC that decays to zero as $r \to \infty$, and it only
exists for a particular pair $(\omega, \gamma)$ near $(\omega_0,
\gamma_0)$. Therefore, both $\omega$ and $\gamma$ must be determined
together with the field.

Inserting expansions (\ref{eq:expE})-(\ref{eq:expk}) into 
Eqs.~(\ref{eq:MaxEq1}) and (\ref{eq:MaxEq2}), and comparing the 
coefficients of $\delta^j$ for $j \geq 1$, we obtain the following 
equation for $\mathbf{E}_j({\bf r})$: 
\begin{eqnarray}
\label{eq:EqEj}
\mathscr{L} \mathbf{E}_j = \gamma_j \mathscr{B} \mathbf{E}_0 + 2 k_0
  k_j \varepsilon_0({\bf r}) \mathbf{E}_0 + \mathbf{F}_j,  
\end{eqnarray}
where 
\begin{eqnarray*}
&& \mathscr{L}  =  (\nabla + i \gamma_0 \mathbf{e}_3) \times (\nabla + i
                  \gamma_0 \mathbf{e}_3) \times \cdot  \ - k_0^2
   \varepsilon_0({\bf r}), \\
&& \mathscr{B}  =  - i \left[  (\nabla + i \gamma_0 \mathbf{e}_3) \times
   \mathbf{e}_3 \times \cdot  + \mathbf{e}_3 \times (\nabla + i \gamma_0
   \mathbf{e}_3) \times \cdot \ \right],  \\
&& \mathbf{F}_1 = s({\bf r}) k_0^2 \mathbf{E}_0, 
\end{eqnarray*}
and $\mathbf{F}_j$, for $j > 1$, are given in Appendix A. Both
$\mathscr{L}$ and $\mathscr{B}$ are differential operators independent
of $j$.  $\mathscr{L}$ is the differential operator associated
 with Eq.~(\ref{eq:MaxEq1}) for  $\omega_0$,  $\gamma_0$ and the unperturbed
 waveguide.  Since ${\bf E}_0$
 satisfies condition  (\ref{eq:RealCondition}), we can verify that
 $\mathscr{B} \mathbf{E}_0$ also satisfies  (\ref{eq:RealCondition}).  
The vector function $\mathbf{F}_j$ depends on $\gamma_m$, $k_m$ and
$\mathbf{E}_m$ for $0\le m < j$. 
For each $j \geq 1$, we need to determine a real $k_j$ and
a real $\gamma_j$, show that $\mathbf{E}_j$ can be solved from
Eq.~(\ref{eq:EqEj}),  and  it decays to zero exponentially as $r \to \infty$ and
satisfies conditions (\ref{eq:RealCondition}) and
(\ref{eq:EvenInX}). We establish the result
  recursively. For the $j$th step, it is assumed that for each $m$ satisfying
  $0\le m   < j$,  we already have a real $k_m$, a real $\gamma_m$, and a
  vector function $\mathbf{E}_m$
  satisfying  (\ref{eq:RealCondition}) and (\ref{eq:EvenInX}).

If such an ${\bf E}_j$ exists, we can show (see Appendix A) that 
\begin{equation}
  \label{twozero}
\int_{\mathbb{R}^2} \overline{\bf E}_0 \cdot  \mathscr{L} {\bf E}_j \, d{\bf 
  r} = \int_{\mathbb{R}^2} \overline{\bf E}^{(s)} \cdot \mathscr{L} {\bf E}_j \,
d{\bf    r} = 0, 
\end{equation}
where ${\bf E}^{(s)}$ is a scattering solution constructed in
Sec.~III (for $\omega_0$, $\gamma_0$, and the unperturbed waveguide)  satisfying conditions (\ref{eq:RealCondition}),
(\ref{eq:EvenInX}) and (\ref{orthogonal}). 
Replacing $\mathscr{L} {\bf E}_j$ by the right hand side of
Eq.~(\ref{eq:EqEj}), we have 
\begin{eqnarray}
\label{eq:systemEq1}
&& \int_{\mathbb{R}^2} \overline{\mathbf{E}}_0 \cdot \left(  \gamma_j
  \mathscr{B} \mathbf{E}_0  + 2 k_0 k_j  \varepsilon_0(\mathbf{r})  \mathbf{E}_0 +
  \mathbf{F}_j  \right)d \mathbf{r} = 0,  \\ 
\label{eq:systemEq2}
&& \int_{\mathbb{R}^2} \overline{\mathbf{E}}^{(s)} \cdot \left(
   \gamma_j \mathscr{B} \mathbf{E}_0  + 2 k_0 k_j  \varepsilon_0(\mathbf{r}) \mathbf{E}_0
   + \mathbf{F}_j  \right)d \mathbf{r} = 0. 
\end{eqnarray}
The above can be written as 
\begin{equation}
\label{eq:system}
\mathbf{A} \left[ \begin{matrix} \gamma_j \\ k_j \end{matrix} \right]
= \left[ \begin{matrix} b_{1j} \\ b_{2j} \end{matrix} \right], \quad
 {\bf A} = \left[ \begin{matrix} a_{11} & a_{12} \\  a_{21} &
    a_{22}  \end{matrix} \right], 
\end{equation}
where 
\begin{eqnarray}
&&   a_{11} = \int_{\mathbb{R}^2} \overline{\mathbf{E}}_0 \cdot \mathscr{B}
             \mathbf{E}_0\,  d \mathbf{r}, \\
  && a_{12} = 2 k_0 \int_{\mathbb{R}^2}  \varepsilon_0(\mathbf{r})
     |\mathbf{E}_0|^2  \, d\mathbf{r}  = 2L^2k_0 > 0,  \\
&& a_{21}  = \int_{\mathbb{R}^2} \overline{\mathbf{E}}^{(s)} \cdot 
   \mathscr{B} \mathbf{E}_0\  d \mathbf{r}, \\
  && a_{22} = 2 k_0 \int_{\mathbb{R}^2}  \varepsilon_0(\mathbf{r})
     \overline{\mathbf{E}}^{(s)}  \cdot  \mathbf{E}_0 \, d\mathbf{r} =0, \\
  && b_{1j} =  -      \int_{\mathbb{R}^2} \overline{\mathbf{E}}_0 \cdot
     \mathbf{F}_j \, d\mathbf{r}, \\ 
  && b_{2j} = -  \int_{\mathbb{R}^2} \overline{\mathbf{E}}^{(s)} \cdot
     \mathbf{F}_j \,  d \mathbf{r}. 
\end{eqnarray}
Since $\mathscr{B} {\bf E}_0$ satisfies condition 
(\ref{eq:RealCondition}), $a_{11}$ and $a_{21}$ are real. In addition, 
$a_{12} > 0$, and $a_{22}=0$ due to the orthogonality condition  
(\ref{orthogonal}). Therefore, all entries of ${\bf A}$ are real.
Meanwhile, since ${\bf F}_j$ depends on $\gamma_m$,
  $k_m$ and ${\bf E}_m$ for $0\le m < j$,  $\gamma_m$ and
$k_m$ are real,  and ${\bf E}_m$ satisfy condition
(\ref{eq:RealCondition}),   $\mathbf{F}_j$ also satisfies
condition  (\ref{eq:RealCondition}), and  thus $b_{1j}$ and $b_{2j}$ are
real. Therefore, when ${\bf A}$ is  invertible, $\gamma_j$ and $k_j$
can be solved, and they are real.

Although Eqs.~(\ref{eq:systemEq1}) and (\ref{eq:systemEq2}) 
are derived assuming Eq.~(\ref{eq:EqEj}) has a solution, the linear system (\ref{eq:system}) does not depend on
${\bf E}_j$. In the following, we show that if $\gamma_j$ and $k_j$
satisfy (\ref{eq:system})  and ${\bf A}$ is
invertible, then
Eq.~(\ref{eq:EqEj}) indeed has a solution ${\bf
  E}_j$ that decays to zero exponentially  as $r \to \infty$ and
satisfies conditions (\ref{eq:EvenInX}) and
(\ref{eq:RealCondition}).

Since ${\bf F}_j$ depends on ${\bf E}_m$ for $0\le m < j$ and ${\bf 
  E}_m \to 0$ exponentially as $r \to \infty$, the right hand side of 
Eq.~(\ref{eq:EqEj}) tends to zero as $r \to \infty$ and can be 
regarded as a source distributed around the waveguide core. Therefore, 
we  require the solutions of Eq.~(\ref{eq:EqEj}), if they exist,  to satisfy an 
outgoing radiation condition. Since only the fundamental TE slab mode
can radiate out power (in the $x$ direction), the outgoing radiation
condition gives  rise to 
\begin{equation}
\label{Ejbigx}
\mathbf{E}_j({\bf r}) \sim  d^{\pm}_j \mathbf{u}_{\pm}({\bf r}), \quad x \to \pm \infty, 
\end{equation}
for some coefficients  $d^{\pm}_j$.
Since $\mathscr{L} \mathbf{E}_0 = 0$, Eq.~(\ref{eq:EqEj}) is a 
singular equation, and it has solutions if and only if the right hand 
side is orthogonal with the kernel of $\mathscr{L}$. Since the BIC is 
non-degenerate, the kernel of $\mathscr{L}$ is the one-dimensional 
vector space spanned by ${\bf E}_0$. Therefore, 
if  condition (\ref{eq:systemEq1}) is satisfied, Eq.~(\ref{eq:EqEj})  has 
solutions satisfying (\ref{Ejbigx}). The solutions are not unique, but 
all solutions of   Eq.~(\ref{eq:EqEj}) must have the same asymptotic
coefficients $d_j^\pm$. This is so, because the difference of two
solutions of Eq.~(\ref{eq:EqEj}) satisfies a homogeneous equation 
and does not radiate out power to infinity. Furthermore, 
since all ${\bf E}_m$, for $0 \le m < j$, satisfy condition (\ref{eq:EvenInX})
and the perturbation profile $s({\bf r})$ satisfies condition
(\ref{eq:Eps_symm}), it is easy to verify that ${\bf E}_j$ also satisfies
condition (\ref{eq:EvenInX}). Therefore, $d_j^+ = d_j^-$.

The second condition, Eq.~(\ref{eq:systemEq2}), can be used to show that 
$d_j^\pm = 0$. 
For any positive $h$, let $\Omega_h$ be the domain given by $|x| < h$ and $|y| < \infty$, then Eqs.~(\ref{eq:EqEj}) and (\ref{eq:systemEq2}) clearly imply that 
\begin{equation}
  \lim\limits_{h \to \infty} \int_{\Omega_h}
  \overline{\mathbf{E}}^{(s)} \cdot \mathscr{L} \mathbf{E}_j \ d 
  \mathbf{r} = 0. 
\end{equation}
In Appendix B, we show that 
\begin{equation}
\label{eq:coeff_E1}
\lim\limits_{h \to +\infty} \int_{\Omega_h}
\overline{\mathbf{E}}^{(s)} \cdot \mathscr{L} \mathbf{E}_j \
d\mathbf{r} =  - 2 i \alpha_0^{\rm te} (d^+_j + d^-_j) C, 
\end{equation}
where $C\ne 0$ is the constant given in
Eq.~(\ref{getreal}). This implies that $d_j^\pm=0$, and thus 
${\bf E}_j \to 0$ exponentially as $r \to \infty$. 

It can be easily verified that the right hand side of
  Eq.~(\ref{eq:EqEj}) satisfies condition
  (\ref{eq:RealCondition}). Thus, for any solution ${\bf E}_j$, we can
  construct a vector field $\tilde{\bf E}_j$ as in Eq.~(\ref{tildeE}),
  then $\tilde{\bf 
  E}_j$ and $({\bf E}_j + \tilde{\bf E}_j)/2$ also satisfy
Eq.~(\ref{eq:EqEj}) and they decay to zero exponentially as $r \to \infty$. Therefore, we
can replace ${\bf E}_j$ by $({\bf E}_j + \tilde{\bf E}_j)/2$ (if
necessary), and assume ${\bf E}_j$ satisfies condition (\ref{eq:RealCondition}).

The above proof is for a BIC satisfying the symmetry condition
(\ref{eq:EvenInX}). If the BIC satisfies condition (\ref{eq:OddInX}),
we need to use the scattering solution that also satisfies (\ref{eq:OddInX}),
then $d_j^+= -d_j^-$, and $d_j^+-d_j^-$ should appear in the right
hand side of Eq.~(\ref{eq:coeff_E1}). If $s({\bf r})$ does not
satisfy condition (\ref{eq:Eps_symm}), the above proof fails,
because we no longer have $d_j^+ = \pm d_j^-$, then $d_j^\pm\ne 0$ in
general, and ${\bf E}_j$ does not decay to zero as $r \to \infty$.
To
ensure that matrix ${\bf A}$ is invertible, the BIC of the unperturbed
waveguide is required to satisfy 
\begin{equation}
  \label{generic}
  \int_{\mathbb{R}^2} \overline{\bf E}^{(s)} \cdot \mathscr{B} {\bf E}_0
  \, d{\bf r} \ne 0, 
\end{equation}
where ${\bf E}^{(s)}$ satisfies the orthogonality condition
(\ref{orthogonal}). 
In  summary, if the original waveguide is symmetric in $x$ and has a non-degenerate BIC satisfying conditions
(\ref{eq:OneGuidedMode}) and (\ref{generic}), and the perturbation
is also symmetric in $x$, then the perturbed waveguide has a BIC with
a slightly different 
frequency and a slightly different propagation constant. 


\section{Numerical examples}

\begin{figure}[htp]
\centering 
\includegraphics[scale=0.7]{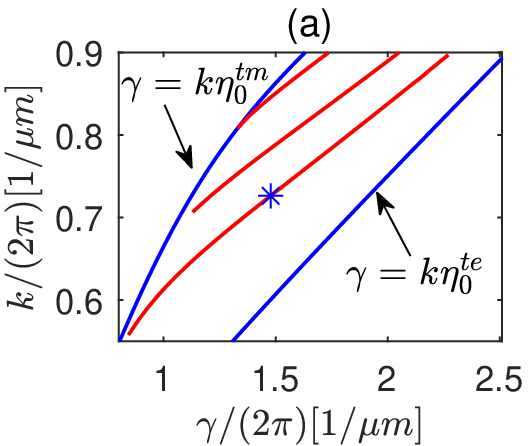}
\includegraphics[scale=0.7]{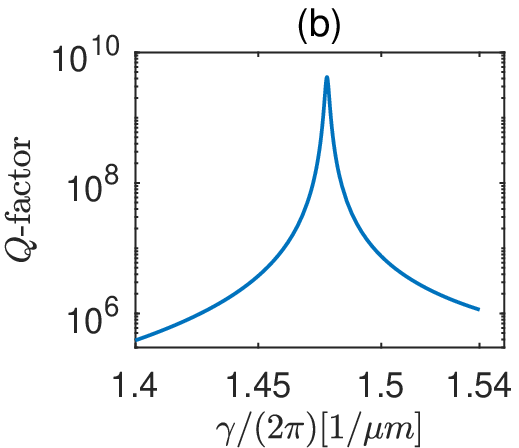}
\includegraphics[scale=0.7]{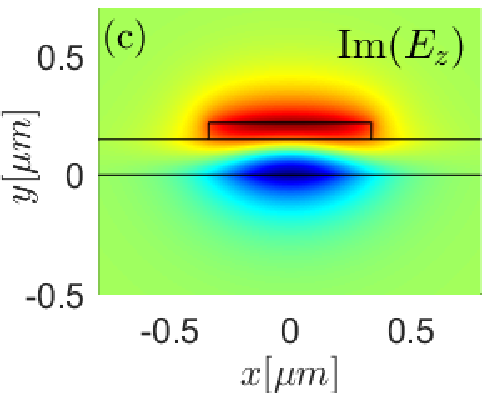}
\includegraphics[scale=0.7]{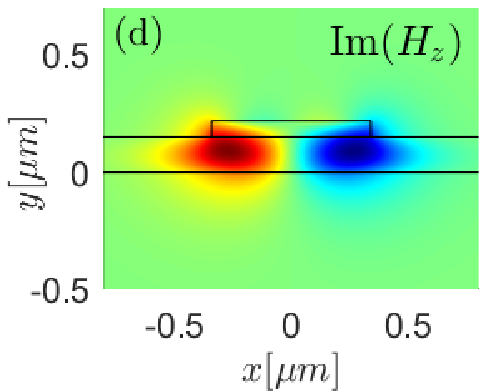}
\includegraphics[scale=0.7]{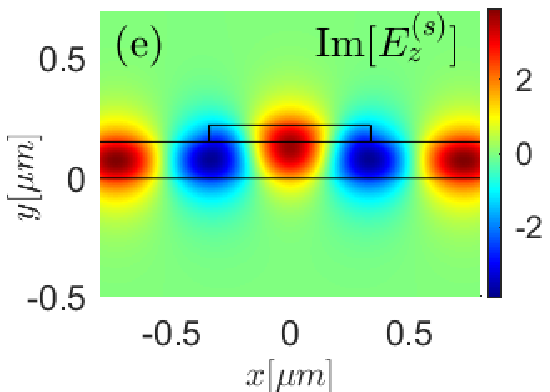}
\includegraphics[scale=0.7]{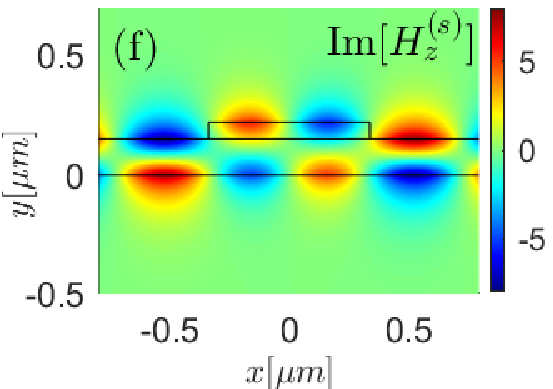}
\caption{A rectangular ridge waveguide with a BIC. (a): dispersion curves of the
  fundamental TE and TM slab modes (solid blue curves) and a few resonant modes (solid red
  curves), and a BIC with $(\gamma_0, k_0) \approx (1.4794, 0.7266) (2 \pi
  / \mu m)$. (b): $Q$ factor of the resonant modes in a band with the BIC. (c) and (d): the imaginary parts of $E_z$ and
  $H_z$ of the BIC, respectively.  (e) and (f): the imaginary
  parts of $E^{(s)}_z$ and $H^{(s)}_z$ of the scattering solution.}
\label{fig:Fig2}
\end{figure}

In this section, we present some numerical examples to demonstrate
the robustness of BICs in optical waveguides. We start with a silicon
ridge waveguide with a rectangular ridge, a SiO$_2$ substrate,  and an
air cladding. The dielectric constants of the cladding, the ridge and
the slab, and the substrate are $\varepsilon_0=1$, $\varepsilon_1 = 12.25$,
and $\varepsilon_2 = 2.1025$, respectively.
The geometric parameters, as shown in Fig.~\ref{fig:structure}, are 
$W = 0.68 \mu m$,  $h_r = 0.07 \mu m$, $h_s = 0.15 \mu m$, and 
$\theta = 0$.  This waveguide has a BIC with freespace wavenumber 
$ k_0  \approx  0.7266 (2\pi/\mu m)$ (i.e. freespace wavelength
$\lambda_0 \approx  1.3762 \mu m$) and  propagation constant $\gamma_0 \approx
1.4794 (2\pi / \mu m)$. In Fig.~\ref{fig:Fig2}(a),
we show the dispersion curves of the fundamental TE and TM slab modes
(solid blue curves) and the dispersion curves of some resonant modes (solid red 
lines, for real part of $k=\omega/c$ only). The BIC, marked as
``$\ast$'' in Fig.~\ref{fig:Fig2}(a),  satisfies condition
(\ref{eq:OneGuidedMode}) and is a special point on a band of
resonant modes. In Fig.~\ref{fig:Fig2}(b), we show the $Q$ factor of
this particular band for $\gamma $ near $\gamma_0$.
The BIC has been scaled to satisfy conditions (\ref{eq:RealCondition})
and (\ref{eq:NormalizationBIC})
with $L=1 \mu m$. 
The $z$-components of the electric field and a scaled magnetic field of the BIC are
shown in Fig.~\ref{fig:Fig2}(c) and (d), respectively. The scaled magnetic
field is obtained by multiplying the free space impedance to the
original magnetic field. Due to the normalization,
Eq.~(\ref{eq:NormalizationBIC}), all field components are
dimensionless. From Fig.~\ref{fig:Fig2}(c), it is clear that the BIC
satisfies symmetry condition (\ref{eq:OddInX}).  Following Sect.~III, we
calculate a scattering solution satisfying conditions
(\ref{eq:RealCondition}), (\ref{eq:OddInX}) and
(\ref{orthogonal}). The $z$ components of 
the scattering solution, i.e. $E^{(s)}_z$ and $H^{(s)}_z$, are shown
in Fig.~\ref{fig:Fig2}(e) and (f).

\begin{figure}[htp]
\centering 
\includegraphics[scale=0.7]{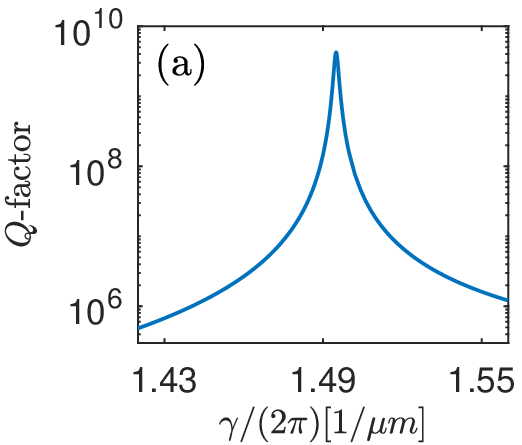}
\includegraphics[scale=0.7]{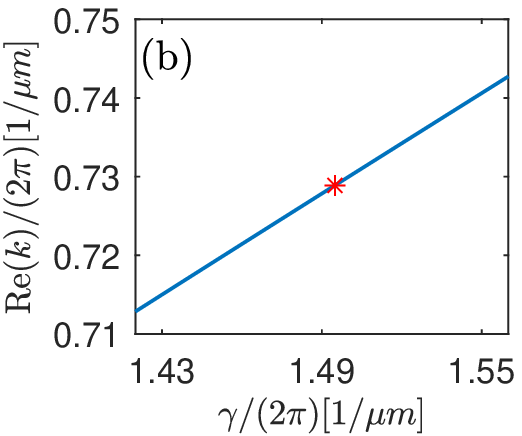}
\includegraphics[scale=0.7]{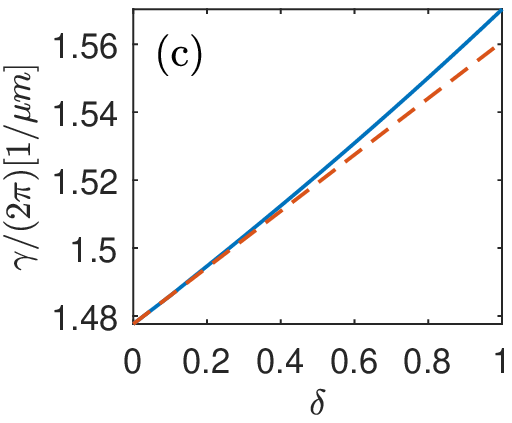}
\includegraphics[scale=0.7]{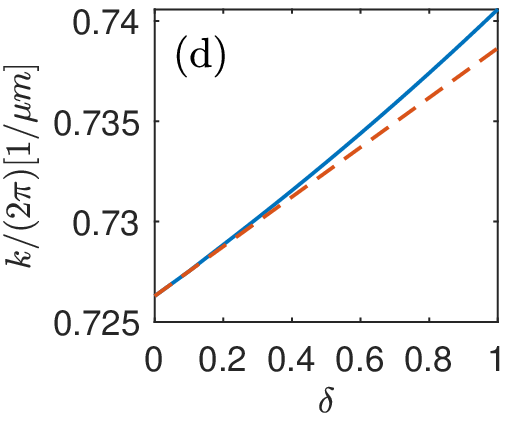}
\caption{BIC in a ridge waveguide with a varying dielectric function. (a):
  $Q$ factor of resonant modes on a band with a BIC, for $\delta = 0.2$. (b):
  Dispersion curve of the same band. The BIC is located at $(\gamma,
  k) \approx  (1.4947, 0.7288) (2 \pi / \mu   m)$ and shown as the
  ``$\ast$''.  (c) and (d): Values of $\gamma$ and $k$ of the BIC for different
  $\delta$ (solid blue curves) and their first order approximations
  (red dashed lines).}
\label{fig:Fig3}
\end{figure}

It is easy to verify that the BIC satisfies condition
(\ref{generic}). Therefore, according to the theory developed in
Sect.~IV, the BIC should be robust with
respect to any perturbation that preserves condition
(\ref{eq:Eps_symm}), i.e., the left-right mirror symmetry. 
To validate the theory, we consider a perturbation profile
\begin{equation}
s({\bf r}) = \sin \left[\frac{ \pi(y - h_s)}{2h_r} \right],  
\end{equation}
for $ |x| < W/2 $ and $ h_s < y < h_s+ h_r$, and $s({\bf r}) = 0$,
otherwise, and study the waveguide numerically for $\delta \in [0,
1]$. The numerical results confirm that the BIC exists continuously
as $\delta$ is increased from $0$ to $1$. The propagation constant
$\gamma$ and freespace wavenumber $k$ of the BIC are shown
as functions of $\delta$ (solid blue curves) in Fig.~\ref{fig:Fig3}(c)
and (d),
respectively. For $\delta = 0.2$, the BIC is obtained with $\gamma \approx 1.4947
(2\pi/\mu m)$ and  $k \approx 0.7288 (2\pi / \mu m)$. 
In Fig.~\ref{fig:Fig3}(a) and (b), we show the $Q$ factor and the
dispersion curve, respectively,  for a band of
resonant modes containing the BIC [shown as the ``$\ast$"  in
Fig.~\ref{fig:Fig3}(b)]. In Fig.~\ref{fig:Fig3}(c) and (d), we also
show first order approximations, 
$\gamma \approx \gamma_0 + \gamma_1 \delta$ and $k \approx k_0 +
k_1 \delta $,  as the red dashed lines, where $\gamma_1$ and $k_1$ are
first order terms used in Sect.~III. Based on the numerical solutions
of the BIC and the scattering solution of the unperturbed waveguide,
we obtain $\gamma_1 \approx 0.5240 (1/\mu m)$ and $k_1 \approx 0.0781
(1/\mu m)$. 

\begin{figure}[htp]
\centering 
\includegraphics[scale=0.7]{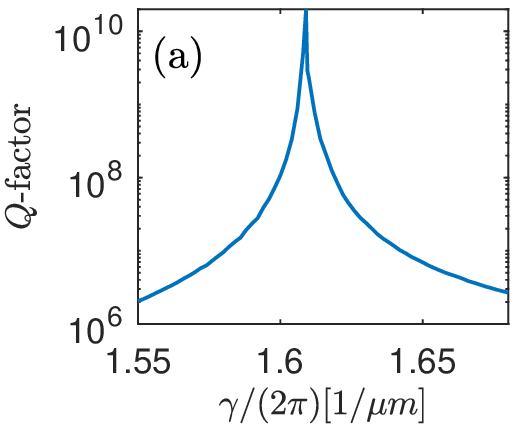}
\includegraphics[scale=0.7]{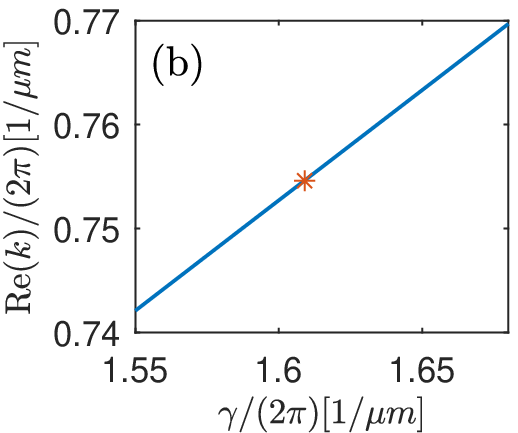}
\includegraphics[scale=0.7]{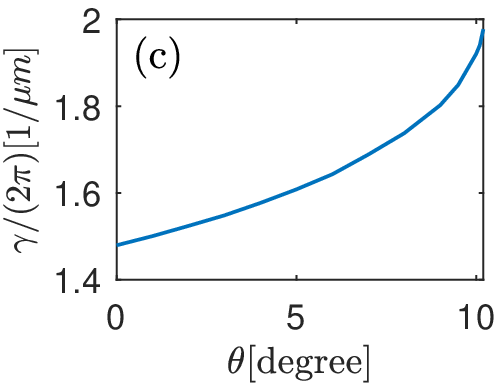}
\includegraphics[scale=0.7]{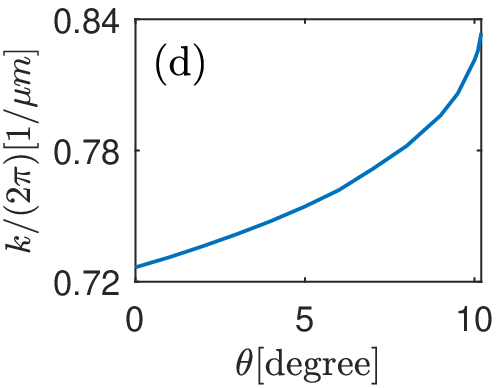}
\includegraphics[scale=0.7]{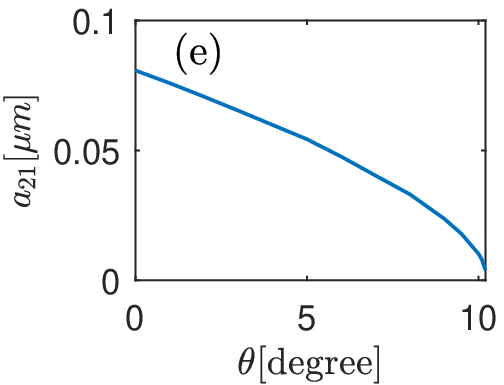}
\caption{BIC in a trapezoidal ridge waveguide. (a): $Q$ factor of
  resonant mode on a band with a BIC, for $\theta=5^\circ$. (b):
  Dispersion curve of the same band. The BIC is located as 
  $(\gamma, k)  \approx (1.6082, 0.7544) (2 \pi / \mu m)$ and shown as
  the ``$\ast$''. (c) and (d): $\gamma$ and $k$ of the BIC for
  different values of $\theta$.  (e): The $(2,1)$ entry of
  matrix ${\bf A}$  as a function of $\theta$.}
\label{fig:Fig4}
\end{figure}

As a second test for the theory, we consider a waveguide
with an isosceles trapezoidal ridge. For the same set of parameters, $\varepsilon_0$,
$\varepsilon_1$, $\varepsilon_2$, $W$, $h_r$, and $h_s$,  we study the
waveguide allowing $\theta$, shown in Fig.~\ref{fig:structure}, to
increase from $0^\circ$ to $10.2^\circ$. The numerical results reveal a BIC that depends
on $\theta$ continuously. The propagation constant $\gamma$ and
freespace wavenumber $k$ of the BIC are shown as functions of $\theta$
in Fig.~\ref{fig:Fig4}(c) and (d), respectively. 
For $\theta = 5^{\circ}$, the
BIC, as usual, is a special point on a band of resonant modes. The $Q$
factor and the dispersion 
curve of this band 
are shown in Fig.~\ref{fig:Fig4}(a) and (b),
respectively.  The BIC is obtained with $(\gamma, k) \approx  (1.6082,
0.7544) (2\pi/\mu m)$ and shown as the ``$\ast$'' in
Fig.~\ref{fig:Fig4}(b).
Additional numerical results indicate that the BIC ceases to exist for $\theta >
10.2^{\circ}$. It appears that the BIC in the waveguide with $\theta
\approx 10.2^\circ$ fails to satisfy condition (\ref{generic}), i.e.,
$a_{21}=0$ and matrix $\mathbf{A}$ is not invertible. In
Fig.~\ref{fig:Fig4}(e), we show $a_{21}$ as a function of $\theta$.
It  indicates  clearly that $a_{12}$ can become zero when $\theta$ is further
increased. 


\section{Conclusion}

A 3D $z$-invariant optical waveguide may have a lateral leakage channel if the
waveguide core is embedded in a background which itself is a 2D
waveguide (typically a slab waveguide). A guided mode of the 3D
waveguide is a BIC if its propagation constant $\gamma$ is smaller
than the propagation constant of the fundamental TE slab mode. In this
paper, we studied BICs satisfying condition (\ref{eq:OneGuidedMode}), so
that only the fundamental TE slab mode can have the same $z$
dependence as the BIC and still propagate in the lateral
direction. For waveguides with the left-right mirror symmetry, i.e.,
condition (\ref{eq:Eps_symm}), we showed that any BIC satisfying
conditions (\ref{eq:OneGuidedMode}) and (\ref{generic}) is robust with
respect to structural perturbations that preserve condition
(\ref{eq:Eps_symm}). The left-right mirror symmetry is imposed so that
the solutions satisfy either condition (\ref{eq:EvenInX}) or condition
(\ref{eq:OddInX}), and the radiation channels to the left and right 
(via the fundamental TE slab mode) are essential the same.
Condition (\ref{generic}) is given assuming the BIC and the scattering solution
are properly scaled, have the same lateral symmetry, and are
orthogonal to each other, i.e., they are required to satisfy
conditions (\ref{eq:RealCondition}),  (\ref{orthogonal}), and (\ref{eq:EvenInX}) or
(\ref{eq:OddInX}).

It is straightforward to extend our result to waveguides without the
left-right mirror symmetry, but with leakage channels in only one
lateral direction, for example, the positive $x$ direction. A guided
mode of such a waveguide is a BIC, if its propagation constant
$\gamma$ is smaller than the propagation constant of the right fundamental
TE  slab mode (of the layered structure for $x>W/2$). If $\gamma$ is larger
than the propagation constants of all other  right slab modes and all
left slab modes (of the layered structure for $x< -W/2$, if they
exist), then the BIC exists in a continuum with only one radiation
channel. Such a BIC, if it is non-degenerate and satisfies the generic
condition (\ref{generic}), is robust with respect to any structure perturbation. The
proof is slightly simpler than the one given in Sect.~III, because we no
longer need to worry about 
the symmetry conditions (\ref{eq:EvenInX}) and (\ref{eq:OddInX}). It
is only necessary to ensure the BIC and the scattering solution satisfy 
conditions (\ref{eq:RealCondition}) and (\ref{orthogonal}).

The BICs studied in this paper are quite different from those in 
periodic structures sandwiched between two homogeneous
media. They have a different number of direction along which the field
is confined, and they exist in radiation continua provided by
different waves. The left-right mirror symmetry  assumed in this 
paper  and the up-down mirror symmetry for biperiodic
structures with propagating BICs \cite{conrob}, serve the same
purpose, that is, to lock together the radiation channels in the
opposite directions. For biperiodic structures, besides the up-down
mirror symmetry, the in-plane inversion
symmetry is also needed to ensure the robustness of propagating BICs
\cite{conrob}. In contrast,  for waveguides with lateral leakage
channels, robustness of BICs does not require any additional
symmetry. This difference is related to our assumption that the
waveguide is $z$-invariant and the perturbation is $z$-independent.

 \section*{Acknowledgements}

 The authors acknowledge support from the Natural Science Foundation
 of Chongqing, China (Grant No. cstc2019jcyj-msxmX0717),  the program for the Chongqing Statistics Postgraduate Supervisor Team (Grant No. yds183002), and the
 Research Grants Council of Hong Kong Special Administrative Region,
 China (Grant No. CityU 11305518). 
 
 \section*{Appendix A}
For Eq.~(\ref{eq:EqEj}), the vector ${\bf F}_j$ in the right hand side
is 
 \begin{eqnarray*}
&&  \mathbf{F}_j = - i \sum\limits_{m=1}^{j-1} \left( \mathbf{e}_3
                  \times \nabla \times \mathbf{E}_{j-m} 
   + \nabla \times \mathbf{e}_3 \times \mathbf{E}_{j-m} \right) \\
   && + \sum\limits_{m=1}^{j-1} \left( \gamma_m \gamma_{j-m} \mathbf{e}_3 \times \mathbf{e}_3 \times \mathbf{E} + k_m k_{j-m} \varepsilon_0 \mathbf{E} \right) \\  
 && + \sum\limits_{n=1}^{j-1} \sum\limits_{m=0}^{n} \left( \gamma_m
    \gamma_{n-m} \mathbf{e}_3 \times \mathbf{e}_3 \times
    \mathbf{E}_{j-n} + k_m k_{n-m} \varepsilon_0 \mathbf{E}_{j-n} \right) \\
      && + s \sum\limits_{n=0}^{j-1} \sum\limits_{m=0}^n k_m k_{n-m} \mathbf{E}_{j-1-n}. 
 \end{eqnarray*}
Since  $\mathscr{L} \mathbf{E}_0 = 0$, we have 
\begin{eqnarray*}
&&  \overline{\bf E}_0 \cdot \mathscr{L} \mathbf{E} _j  =
  \overline{\bf E}_0 \cdot \mathscr{L} \mathbf{E} _j  -  \mathbf{E} _j
   \cdot \overline{\mathscr{L}}  \overline{\bf E}_0   \\ 
 && = \overline{\bf E}_0  \cdot  \left( \nabla  \times  \nabla
   \times \mathbf{E} _j \right) + i \gamma_0  \overline{\bf E}_0
   \cdot \left[    \nabla  \times \left( \mathbf{e}_3  \times \mathbf{E} _j
             \right) \right]  \\
  && + i \gamma_0  \overline{\bf E}_0 \cdot
   \left[   \mathbf{e}_3   \times \left(   \nabla\times \mathbf{E} _j \right)
   \right] - \gamma_0^2 \overline{\bf E}_0 \cdot \left[   \mathbf{e}_3   \times
   \left( \mathbf{e}_3 \times \mathbf{E} _j \right) \right]  \\
&& -   {\mathbf{E} _j} \cdot  \left( \nabla  \times  \nabla \times
  \overline{\bf E}_) \right)  + i \gamma_0 \mathbf{E} _j \cdot
  \left[    \nabla  \times \left( \mathbf{e}_3  \times \overline{\bf E}_0
  \right) \right]  \\
 && 
+ i \gamma_0  \mathbf{E} _j \cdot \left[
  \mathbf{e}_3   \times \left(   \nabla \times \overline{\bf E}_0 \right)
  \right]   + \gamma_0^2  \mathbf{E} _j \cdot \left[   \mathbf{e}_3
  \times \left( \mathbf{e}_3 \times \overline{\bf E}_0 \right) \right].  
 \end{eqnarray*}
Using the vector identities
\begin{eqnarray*}
&&  {\bf a} \cdot \left( \nabla \times {\bf b} \right) =  {\bf  b} \cdot \left( \nabla
   \times {\bf a} \right)  + \nabla \cdot \left(  {\bf b} \times {\bf a} \right), \\
&&  {\bf a} \cdot \left(  {\bf b} \times {\bf c} \right) = {\bf b}
   \cdot \left( {\bf c} \times {\bf a}
   \right)  = {\bf c} \cdot \left( {\bf a} \times {\bf b} \right), 
\end{eqnarray*}
we can simplify the above as 
\begin{equation}
  \label{FourG}
 \overline{\bf E}_0 \cdot \mathscr{L} \mathbf{E} _j =
 \nabla \cdot {\bf J}, 
\end{equation}
where
\begin{eqnarray*}
&&  {\bf J} =      \left(\nabla \times \mathbf{E} _j \right) \times 
   \overline{\bf E}_0 +   i \gamma_0   \left(\mathbf{e}_3 \times  \mathbf{E} _j \right) 
   \times \overline{\bf E}_0 \cr
&&   -  \left(\nabla \times \overline{\bf E}_0 \right) \times 
   \mathbf{E} _j 
   + i \gamma_0       \left( \mathbf{e}_3 \times \overline{\bf E}_0 \right) 
   \times \mathbf{E} _j. 
\end{eqnarray*}
Integrating both sides of Eq.~(\ref{FourG}) on a disk of radius $a$
(denoted as $D_a$) and using Gauss's Law, we obtain 
\begin{eqnarray*}
&& \int_{D_a} \overline{\bf E}_0 \cdot \mathscr{L} \mathbf{E}_j \ d
   {\bf r} = \int_{\Gamma_a}   {\bf J} \cdot d\mathbf{s}, 
\end{eqnarray*}
where $\Gamma_a$ is the circle of radius $a$. Since $\mathbf{E}_0 \to
0$ exponentially as $r \to \infty$, the line integral in the right
hand side above tends to zero as $a \to \infty$. Therefore
$  \int_{\mathbb{R}^2} \overline{\bf E}_0 \cdot \mathscr{L} \mathbf{E}
  _j \, d {\bf r} = 0$. 
Since we assume $\mathbf{E} _j \to 0$ exponentially as $r \to \infty$,
$\overline{\bf E}^{(s)}\cdot \mathscr{L} \mathbf{E} _j$  
is also integrable on $\mathbb{R}^2$ and the 
integral is also zero.

\section*{Appendix B}
Unlike the case considered in Appendix A,  $\mathbf{E} _j$ is
outgoing as $x \to \pm \infty$, and thus bounded at infinity,
Since $\mathscr{L} \mathbf{E}^{(s)} = 0$, we have 
\[
  \overline{\mathbf{E} }^{(s)}\cdot \mathscr{L} \mathbf{E} _j 
 =  \overline{\mathbf{E} }^{(s)}\cdot \mathscr{L} \mathbf{E} _j -
 \mathbf{E} _j \cdot  \overline{\mathscr{L}} \overline{\mathbf{E} }^{(s)}
= \nabla \cdot {\bf G},
\]
where 
\begin{eqnarray*}
{\bf G} &=& (\nabla \times \mathbf{E} _j) \times 
    \overline{\mathbf{E} }^{(s)} - (\nabla \times \overline{\mathbf{E} }^{(s)} ) \times 
            {\bf E}_j \\
&+&  i \gamma ( \mathbf{e}_3 \times \mathbf{E} _j ) \times 
   \overline{\mathbf{E} }^{(s)} + i \gamma (\mathbf{e}_3 \times
 \overline{\mathbf{E} }^{(s)}) \times    \mathbf{E} _j. 
\end{eqnarray*}
Since $\mathbf{E}_j$ and $\overline{\mathbf{E}}^{(s)}$ decay to zero 
exponentially as $y \to \pm \infty$, we can integrate 
$\overline{\mathbf{E} }^{(s)}\cdot \mathscr{L} \mathbf{E} _j$ on
$\Omega_h$, apply Gauss' Law, and obtain 
\begin{eqnarray*}
 \int_{\Omega_h} \overline{\mathbf{E}}^{(s)} \cdot \mathscr{L}
    \mathbf{E}_1 \ d {\bf r} 
  = \int_{-\infty}^{+\infty}  \mathbf{e}_1 \cdot  \left( {\bf G}|_{x=h} - {\bf G}|_{x=-h} \right)\,  dy, 
\end{eqnarray*}
where   $\mathbf{e}_1 = (1, 0, 0)$ is the unit vector along the $x$
axis.  Based in the asymptotic formulae (\ref{getreal}) and
(\ref{Ejbigx}), it is easy to verify that
\[ 
 \lim\limits_{h \to +\infty} \int_{\Omega_h}
 \overline{\mathbf{E}}^{(s)} \cdot \mathscr{L} \mathbf{E}_j d {\bf r}
 = - 2 i \alpha_0^{\rm te} (d^+_j + d^-_j) C.
\]
That is Eq.~(\ref{eq:coeff_E1}).


\begin{thebibliography}{99}

\bibitem{neumann29} J. von Neumann and E. Wigner, 
 ``\"{U}ber   merkw\"{u}rdige diskrete Eigenwerte,'' 
Phys. Z.  {\bf 30},  465-467 (1929).   

\bibitem{hsu16} C. W. Hsu, B. Zhen, A. D. Stone, 
  J. D. Joannopoulos, and M. Solja\v{c}i\'{c}, 
 ``Bound states in the continuum,'' 
 Nat. Rev. Mater. {\bf 1}, 16048 (2016).

\bibitem{kosh19} K. Koshelev, G. Favraud, A. Bogdanov, Y. Kivshar, and A. Fratalocchi, ``Nonradiating photonics with resonant dielectric nanostructures,''
  Nanophotonics {\bf  8}, 725-745 (2019).

  \bibitem{evans94}  D. V. Evans, M. Levitin and  D. Vassiliev,
   ``Existence theorems for trapped modes,''
 J. Fluid Mech. {\bf 261}, 21-31 (1994).
  
 
\bibitem{mari08} D. C. Marinica, A. G. Borisov, and 
  S. V. Shabanov, ``Bound states in the continuum in photonics,'' 
  \prl\ {\bf 100}, 183902 (2008).   

\bibitem{lee12} J. Lee, B. Zhen, S. L. Chua, W. Qiu, J. D. Joannopoulos, 
  M. Solja\v{c}i\'{c}, and O. Shapira, ``Observation and 
  differentiation of unique high-Q optical resonances near zero wave 
  vector in macroscopic photonic crystal slabs,'' \prl\ {\bf 109}, 
  067401 (2012). 
 
\bibitem{hsu13_2} C. W. Hsu, B. Zhen, J. Lee, S.-L. Chua, 
  S. G. Johnson, J. D. Joannopoulos, and M. Solja\v{c}i\'{c}, 
  ``Observation of trapped light within the radiation continuum,'' 
  Nature {\bf 499}, 188--191 (2013). 

\bibitem{bulg14b} E. N. Bulgakov and A. F. Sadreev, ``Bloch 
  bound states in the radiation continuum in a periodic array of 
  dielectric rods,''   \pra\ {\bf 90}, 053801 (2014). 





\bibitem{bonnet97} A.-S. Bonnet-Ben Dhia and F. Mah\'{e},
  ``A guided mode in the range of the radiation modes for a rib waveguide,''
  J. Opt. {\bf 28}, 41-43 (1997).


 \bibitem{plot11} Y. Plotnik, O. Peleg, F. Dreisow, M. Heinrich, 
  S. Nolte, A. Szameit, and M. Segev, 
 ``Experimental observation of optical bound states in the 
 continuum,'' 
 \prl\ {\bf 107}, 183901 (2011). 

\bibitem{weim13} S. Weimann, Y. Xu, R. Keil, A. E. Miroshnichenko, 
A. T\"{u}nnermann, S. Nolte, A. A. Sukhorukov, A. Szameit, and Y. S. Kivshar, 
 ``Compact surface Fano states embedded in the continuum of the 
 waveguide arrays,''  
 \prl\ {\bf 111}, 240403 (2013). 

  
\bibitem{zou15}   C.-L. Zou, J.-M. Cui, F.-W. Sun, X. Xiong, X.-B. Zou, Z.- F. Han, and G.-C. Guo, ``Guiding light through optical bound states in the continuum for ultrahigh-$Q$ microresonantors,''
  Laser Photonics Rev.   {\bf 9}, 114-119  (2015).

\bibitem{hope16} A. P. Hope, T. G. Nguyen, A. Mitchell, and W. Bogaerts,
  ``Quantitative analysis of TM lateral leakage in foundry fabricated silicon rib waveguides,''
  \ptl\ {\bf 28}, 493-496 (2016).

\bibitem{bezus18} E. A. Bezus, D. A. Bykov, and L. L. Doskolovich,
  ``Bound states in the continuum and high-$Q$ resonances supported by a dielectric  ridge on a slab waveguide,''
  Photonics Research  {\bf  6}, 1084-1093 (2018).
  
\bibitem{nguyen19} T. G. Nguyen, G. Ren, S. Schoenhardt, M. Knoerzer, A. Boes, and A. Mitchell,
  ``Ridge resonance in silicon photonics harnessing bound states in the continuum,''
  Laser Photonics Rev. {\bf 13},  1900035 (2019).

\bibitem{yu19a} Z. Yu, Y. Wang, B. Sun, Y. Tong, J.-B. Xu, H.  K. Tsang, and X. Sun,
  ``Hybrid 2D-material photonics with bound states in the continuum,''
  Adv. Opt. Mater.   {\bf 7}, 1901306 (2019).

\bibitem{yu19} Z. Yu, X. Xi, J. Ma,   H. K. Tsang,  C.-L. Zou, and X. Sun,
  ``Photonic integrated circuits with bound states in the continuum,''
  Optica  {\bf 6}, 1342-1348 (2019).

\bibitem{byk20} D. A. Bykov, E.  A. Bezus, and L. L. Doskolovich,
  ``Bound states in the continuum and strong phase resonances in 
  integrated Gires-Tournois interferometer,'' 
  Nanophotonics {\bf 9}(1), 83-92 (2020).

\bibitem{yu20}   Z. Yu, Y. Tong, H. K. Tsang, and X. Sun,
``High-dimensional communication on etchless lithium niobate platform with photonic
bound states in the continuum,''
Nature Communications {\bf 11}, 2602 (2020). 

  

  \bibitem{bulg17prl} E. N. Bulgakov and D. N. Maksimov, ``Topological 
   bound states in the continuum in arrays of dielectric spheres,'' 
 \prl\ {\bf 118}, 267401 (2017). 

\bibitem{sadbel19} Z. F. Sadrieva, M. A. Belyakov, M. A. Balezin, P. 
  V. Kapitanova, E. A. Nenasheva,  A. F. Sadreev,  and A. A. Bogdanov,
``Experimental observation of a symmetry-protected bound state in the 
continuum in a chain of dielectric disks,'' 
\pra\  {\bf 99}, 053804 (2019). 

 \bibitem{gomis17} J. Gomis-Bresco, D. Artigas,  and L. Torner, 
  ``Anisotropy-induced photonic bound states in the continuum,'' 
  Nature Photonics  {\bf 11}, 232--237 (2017).


  \bibitem{kosh18} K. Koshelev, S. Lepeshov, M. Liu, A. Bogdanov, and 
  Y. Kivshar, ``Asymmetric metasurfaces with high-$Q$
    resonances governed by bound states in the contonuum,'' 
    \prl\, {\bf     121}, 193903 (2018). 

  \bibitem{lijun18pra} L.  Yuan and  Y. Y.  Lu, 
  ``Bound states in the continuum on periodic structures surrounded by 
  strong resonances,'' 
\pra\ {\bf 97}, 043828 (2018). 
    
\bibitem{hu18} Z. Hu and Y. Y. Lu, ``Resonances and bound states in 
  the continuum on periodic arrays of slightly noncircular 
  cylinders,'' J. Phys. B: At. Mol. Opt. Phys.  {\bf 51}, 035402 (2018). 

\bibitem{zhen20b} Z. Hu, L. Yuan, and Y. Y. Lu, ``Resonant field 
  enhancement near bound states in the continuum on periodic 
  structures,'' 
\pra\
{\bf 101}, 043825 (2020) 

\bibitem{lijun20} L.  Yuan and Y. Y.  Lu, ``Perturbation theories 
  for symmetry-protected bound states in the continuum on 
  two-dimensional periodic structures,'' \pra\
  {\bf 101}, 043827 (2020).


\bibitem{zhen14} B. Zhen, C. W. Hsu, L. Lu, A. D. Stone, and M. 
Solja\v{c}i\v{c},  ``Topological nature of optical bound 
states in the continuum,'' 
\prl\ {\bf 113}, 257401 (2014).

\bibitem{bulg17pra} E. N. Bulgakov and D. N. Maksimov, 
``Bound states in the continuum and polarization singularities in 
periodic arrays of dielectric rods,'' 
\pra\ {\bf 96}, 063833 (2017). 

\bibitem{yuan17_4} L. Yuan and Y. Y. Lu,  ``Bound states in the 
  continuum on periodic structures: perturbation theory and 
  robustness,'' \ol\ {\bf 42}(21) 4490-4493 (2017). 

\bibitem{yuan20b} L. Yuan and Y. Y. Lu, ``Parametric dependence of 
    bound states in the continuum on periodic structures,'' 
    \pra\ {\bf 102}, 033513  (2020). 
  
\bibitem{conrob} L. Yuan and Y. Y. Lu, ``Conditional robustness of 
  propagating bound states in the continuum on biperiodic 
  structures,'' 
  arXiv preprint arXiv:2001.00832  (2020).


\bibitem{marcuse} D. Marcuse, Theory of Dielectric Optical Waveguides, 
  2nd ed. (Academic Press, Boston, 1991). 

  
\end{thebibliography}
\end{document}